\documentstyle [12pt] {article}

\catcode`@=12
\begin{document}
\thispagestyle{empty}
\begin{flushright} UCRHEP-T145\\TIFR/TH/95-17\\April 1995\
\end{flushright}
\vspace{0.5in}
\begin{center}
{\Large \bf Model of Four Light Neutrinos\\
in the Light of All Present Data\\}
\vspace{1.5in}
{\bf Ernest Ma\\}
{\sl Department of Physics, University of California, Riverside,
California 92521\\}
\vspace{0.1in}
{\bf Probir Roy\\}
{\sl Tata Institute of Fundamental Research, Homi Bhabha Road, Bombay
400005, India\\}
\vspace{1.5in}
\end{center}
\begin{abstract}\
Motivated by existing and recent data on possible neutrino oscillations,
we propose a model of four light neutrinos ($\nu_e, \nu_\mu, \nu_\tau$, and
a singlet $\nu_S$) with a pattern of masses and mixing derivable from
a discrete $Z_5$ symmetry and the seesaw mechanism.  Atmospheric neutrino
oscillations occur between $\nu_\mu$ and $\nu_\tau$ as pseudo-Dirac partners;
whereas solar neutrino oscillations occur between
$\nu_e$ and $\nu_S$, a linear combination of which is massless.  Additional
oscillations may occur between $\nu_e$ and $\nu_\mu$ to account for the
recent observation of the LSND (Liquid Scintillator Neutrino Detector)
experiment.
\end{abstract}

\newpage
\baselineskip 24pt

The physics of light neutrinos appears poised on the edge of major
discoveries.  Many hints have accumulated over the past years towards
nonzero masses and flavor mixing of these special elementary fermions
without charge.  They include the solar neutrino deficit\cite{1}, the
atmospheric neutrino anomaly\cite{2}, the need for a cosmological hot
dark matter component\cite{3}, and finally the excess of $\bar \nu_\mu
\rightarrow \bar \nu_e$ events observed recently by the LSND (Liquid
Scintillator Neutrino Detector) experiment\cite{4}.  If all these strands
are put together in a total picture, along with the nonobservations of
neutrinoless double $\beta$-decay\cite{5} and any depletion of reactor
antineutrinos (such as from the Bugey experiment\cite{6}), one is led
naturally to a
variation of a simple scenario already proposed\cite{7}.  We offer in this
paper a theoretical understanding of this specific scenario in terms of
a hierarchical seesaw model of the masses and flavor mixing of four light
neutrinos with an assumed discrete $Z_5$ symmetry.

We recount the constraints from the above observations.  In a two-oscillator
picture involving $\nu_e$, the solar neutrino deficit favors either of two
matter-enhanced oscillation solutions with $\delta m^2_{e\alpha} \sim 10^{-6}$
to $10^{-5}$ eV$^2$, $\sin^2 2 \theta_{e\alpha} \sim 5 \times 10^{-3}$ and
with $\delta m^2_{e\alpha} \sim 10^{-6}$ to $10^{-4}$ eV$^2$, $\sin^2
2 \theta_{e\alpha} > 0.4$ or a vacuum oscillation solution with $\delta
m^2_{e\alpha} \sim 10^{-10}$ eV$^2$, $\sin^2 2 \theta_{e\alpha} > 0.75$.
Here $\delta m^2_{e\alpha}$ is the difference of the squares of the two
neutrino masses and $\theta_{e\alpha}$ is the mixing angle between $\nu_e$
and another light neutrino $\nu_\alpha$.  The atmospheric neutrino
anomaly can be understood in a two-oscillator scenario involving
$\nu_\mu$ and another neutrino $\nu_\beta$ with $\delta m^2_{\mu\beta} \sim
10^{-2}$ eV$^2$ and $\sin^2 2 \theta_{\mu\beta} = O(1)$.  The LSND results,
on the other hand, suggest $\delta m^2_{e\mu} \sim 0.5 - 10$ eV$^2$ and
$\sin^2 2 \theta_{e\mu} \sim (6 \pm 3) \times 10^{-3}$ once again in a
two-oscillator picture.  It is clear that with only the three known
neutrino flavors ($\nu_e, \nu_\mu, \nu_\tau$), we cannot explain all three
$\delta m^2$'s.  The orders of magnitude of the latter are too disparate
to be explained even by invoking three-flavor oscillations.  We are thus
prompted by the need to explain all of the above to add a singlet
neutrino $\nu_S$ which is by itself noninteracting, but will be allowed
to mix with the other neutrinos.  In this way, we are also able to consider
the cosmologically desirable requirement\cite{3} that the masses of the usual
three neutrinos sum up to about 5 eV.

With four neutrinos as suggested above, there is a simple hierarchical
situation involving only two-oscillator scenarios\cite{8}.  Let $\nu_\mu$ and
$\nu_\tau$ be nearly degenerate with masses of about 2.5 eV each and close
to maximal mixing.  The $\delta m^2$ here is about 10$^{-2}$ eV$^2$.  The
singlet neutrino $\nu_S$ has a mass of order $10^{-3}$ eV whereas $\nu_e$
is very much lighter but mixes with $\nu_S$ as well as $\nu_\mu$ by small
amounts.  In this scenario, because $\nu_e$ is lighter than $\nu_\mu$,
there is a potential conflict with rapid neutron capture (r-process) in
Type II supernovae\cite{9}.  However, if the hierarchy is inverted\cite{10}
to avoid this problem, then $\nu_e$ is a few eV in mass and the constraint of
neutrinoless double $\beta$-decay that $m_{\nu_e} < 0.68$ eV\cite{11} cannot
be satisfied.  Given the manifold uncertainties of the r-process calculation
in a hot-bubble scenario, we choose to disregard it in favor of the
double $\beta$-decay constraint.

We view the large mixing but small mass splitting of $\nu_\mu$ and
$\nu_\tau$ as suggestive of their pseudo-Dirac origin.  We also envisage
a massless $\nu_e$ and a very light $\nu_S$ as two Majorana neutrinos with
a small mixing.  A unified seesaw model of these hierarchical masses would
need a Dirac seesaw\cite{12} for the former pair with a minimum $4 \times 4$
matrix and a Majorana seesaw\cite{13} for the latter pair with a minimum
$3 \times 3$ matrix.  An additional small mixing between these two sectors
is necessary for understanding the LSND results and can only come from
both matrices being submatrices of one $7 \times 7$ neutrino mass matrix.

It may appear difficult at first sight to construct a model of the lepton
sector generating naturally a mass matrix with the above requirements from
a symmetry.  However, as shown below, a reasonably simple model does emerge
if one supplements the standard $SU(2)_L \times U(1)_Y$ electroweak gauge
symmetry with a discrete $Z_5$ symmetry\cite{14} which might be the product
of a more fundamental underlying theory.  This discrete symmetry will be
broken spontaneously resulting in the appearance of domain walls.  However,
it is now known\cite{15} that higher-dimensional operators, induced at the
Planck scale, can make such domain walls collapse very quickly after
formation so that we need not consider this as a potential problem.

We take the $Z_5$ elements to be $\omega^{-2}$, $\omega^{-1}$, 1, $\omega$,
and $\omega^2$ with $\omega^5 = 1$.  Let the three lepton families of
left-handed electroweak doublets be denoted by $(\nu_\alpha, l_\alpha)_L$,
with $\alpha = e, \mu, \tau$.  Let the three right-handed charged lepton
singlets $l_{\alpha R}$ be accompanied by four right-handed neutrino singlets
$(\nu_{\alpha R}, \nu_{S R})$.  The $Z_5$ transformations of the leptons
bearing subscripts $e, \mu, \tau, S$ are chosen to be 1, $\omega^{-2}$,
$\omega^2$, $\omega^{-1}$ respectively.  The scalar sector is assumed to
consist of two doublets $\Phi_1 = (\phi^+_1, \phi^0_1)$, $\Phi_2 =
(\phi^+_2, \phi^0_2)$ and
a complex singlet $\chi^0$ transforming as 1, $\omega^{-2}$, and $\omega$
respectively.  The charged lepton mass matrix linking $\bar l_{\alpha L}$
and $l_{\beta R}$ is now of the form
\begin{equation}
{\cal M}_l = \left[ \begin{array} {c@{\quad}c@{\quad}c} a & 0 & d \\
e & b & 0 \\ 0 & 0 & c \end{array} \right],
\end{equation}
where the diagonal entries $a, b, c$ come from the nonzero vacuum expectation
value of $\phi_1^0$ and the off-diagonal entries $d, e$ come from that of
$\phi_2^0$.  The zeros of this matrix are protected at tree level by the
assumed discrete $Z_5$ symmetry.  As it stands, this mass matrix shows
possible $e_L - \tau_L$ but very little $e_L - \mu_L$ mixing.  There could
be substantial $e_R - \mu_R$ mixing, but that is not observable as far as
vector gauge interactions are concerned.

Turning to the neutrino sector, we find the mass matrix spanning
$\bar \nu_{e L}$, $\bar \nu_{\mu L}$, $\bar \nu_{\tau L}$, $\nu_{e R}$,
$\nu_{\mu R}$, $\nu_{\tau R}$, and $\nu_{S R}$ to be given by
\begin{equation}
{\cal M}_\nu = \left[ \begin{array} {c@{\quad}c@{\quad}c@{\quad}c@{\quad}c
@{\quad}c@{\quad}c} 0 & 0 & 0 & m_1 & m_6 & 0 & 0 \\ 0 & 0 & 0 & 0 & m_2 &
0 & 0 \\ 0 & 0 & 0 & m_7 & 0 & m_3 & 0 \\ m_1 & 0 & m_7 & M_1 & 0 & 0 & m_4 \\
m_6 & m_2 & 0 & 0 & m_8 & M_2 & 0 \\ 0 & 0 & m_3 & 0 & M_2 & m_9 & m_5 \\
0 & 0 & 0 & m_4 & 0 & m_5 & 0 \end{array} \right],
\end{equation}
where $m_1, m_2, m_3$ come from $\langle \bar \phi_1^0 \rangle$, $m_6, m_7$
from $\langle \bar \phi_2^0 \rangle$, $m_4, m_9$ from $\langle \chi^0
\rangle$, $m_5, m_8$ from $\langle \bar \chi^0 \rangle$, and $M_1, M_2$ are
allowed mass terms even in the absence of symmetry breaking.  The zeros are
again protected at tree level by the assumed discrete $Z_5$ symmetry.

Note first that ${\cal M}_\nu$ has one zero mass eigenvalue, corresponding
to an eigenstate which is mostly $\nu_e$ as we will show.  In the absence of
symmetry breaking, all the $m$'s are zero and we have only $M_1$ and $M_2$,
corresponding to one massive Majorana fermion and one massive Dirac fermion
respectively.  These allowed masses can be very heavy and will act as the
anchors for the seesaw mechanisms which generate the requisite small neutrino
masses of our model.  Note that the $4 \times 4$ submatrix spanning
$\bar \nu_{e L}$, $\bar \nu_{\mu L}$, $\bar \nu_{\tau L}$, and $\nu_{S R}$
is indeed identically zero, even in the presence of symmetry breaking.

Consider now the situation where only $m_2, m_3$ are made nonzero in addition
to $M_1, M_2$.  Then we have a Dirac seesaw mass $m_2 m_3 / M_2$ for
$\nu_\mu$ and $\nu_\tau$, whereas $\nu_e$ and $\nu_S$ remain massless.
Thus ${\cal M}_\nu$ has a global vector $U(1)$ symmetry as well as two chiral
$U(1)$ symmetries in this case.  If the other $m$'s are also made nonzero,
then these symmetries are broken except of course for that corresponding to
the zero mass eigenvalue noted before.  Hence it is natural\cite{16} to assume
that these other $m$'s are much smaller in magnitude than $m_2, m_3$ which
are in turn much smaller than $M_1, M_2$.  As it turns out, we also need
to make the ratio $m_1/m_4$ small because it corresponds to $\nu_e - \nu_S$
mixing for explaining solar neutrino data.  To summarize, we assume
\begin{equation}
|m_1| << |m_{4,5,6,7,8,9}| << |m_{2,3}| << |M_{1,2}|.
\end{equation}

For very large $M_1, M_2$, the seesaw reduction of ${\cal M}_\nu$ yields
the following $4 \times 4$ mass matrix spanning $\nu_{S R}$, $\bar \nu_{e L}$,
$\bar \nu_{\mu L}$, and $\bar \nu_{\tau L}$:
\begin{equation}
{\cal M}'_\nu = -{1 \over M_1} \left[ \begin{array}
{c@{\quad}c@{\quad}c@{\quad}c} m_4^2 & m_1 m_4 & 0 & m_4 m_7 \\
m_1 m_4 & m_1^2 & 0 & m_1 m_7 \\ 0 & 0 & 0 & 0 \\ m_4 m_7 & m_1 m_7 & 0 &
m_7^2 \end{array} \right] - {1 \over M_2} \left[ \begin{array}
{c@{\quad}c@{\quad}c@{\quad}c} 0 & m_5 m_6 & m_2 m_5 & 0 \\ m_5 m_6 & 0 &
0 & m_3 m_6 \\ m_2 m_5 & 0 & 0 & m_2 m_3 \\ 0 & m_3 m_6 & m_2 m_3 & 0
\end{array} \right].
\end{equation}
Taking the $m_2 m_3/M_2$ entries in the above mass matrix to be
dominant, it is easily seen that a further seesaw reduction yields a
$2 \times 2$ matrix spanning only $\nu_{SR}$ and $\bar \nu_{eL}$ with
elements given by the upper-left-corner submatrix proportional to $1/M_1$.
To find the eigenvalues $\lambda'$ of ${\cal M}'_\nu$ which we denote by
$-\lambda/M_1 M_2$, we write down its characteristic equation:
\begin{eqnarray}
0 &=& \lambda \{ \lambda^3 - \lambda^2 M_2 (m_1^2 + m_4^2 + m_7^2) \nonumber
\\ &~& ~~~ - \lambda [ M_1^2 (m_2^2 m_3^2 + m_2^2 m_5^2 + m_3^2 m _6^2 +
m_5^2 m_6^2) + 2 M_1 M_2 m_1 m_6 (m_4 m_5 + m_3 m_7)] \nonumber \\
&~& ~~~ + M_1^2 M_2 [ m_1^2 m_2^2 (m_3^2 + m_5^2) + (m_3 m_4 - m_5 m_7)^2
(m_2^2 + m_6^2)]\}.
\end{eqnarray}
Using the mass hierarchy assumed in Eq.~(3), the four eigenvalues are
easily obtained:
\begin{eqnarray}
\lambda'_1 &=& 0, \\
\lambda'_2 &=& -{m_4^2 \over M_1}, \\
\lambda'_3 &=& {{m_2 m_3} \over M_2} - {m_7^2 \over {2 M_1}} + {1 \over
{2 M_2}} \left( {{m_2 m_5^2} \over m_3} + {{m_3 m_6^2} \over m_2} \right), \\
\lambda'_4 &=& -{{m_2 m_3} \over M_2} - {m_7^2 \over {2 M_1}} - {1 \over
{2 M_2}} \left( {{m_2 m_5^2} \over m_3} + {{m_3 m_6^2} \over m_2} \right).
\end{eqnarray}
The corresponding mass eigenstates are then related to the interaction
eigenstates by
\begin{equation}
\left[ \begin{array} {c} \nu^c_S \\ \nu_e \\ \nu_\mu \\ \nu_\tau \end{array}
\right] = \left[ \begin{array} {c@{\quad}c@{\quad}c@{\quad}c} -m_1/m_4 &
1 & m_5/m_3 \sqrt 2 & m_5/m_3 \sqrt 2 \\ 1 & m_1/m_4 & -m_6/m_2 \sqrt 2
& m_6/m_2 \sqrt 2 \\ -m_6/m_2 & 0 & -1/\sqrt 2 & 1/\sqrt 2 \\ 0
& -m_5/m_3 & 1/\sqrt 2 & 1/\sqrt 2  \end{array} \right] \left[
\begin{array} {c} \nu_1 \\ \nu_2 \\ \nu_3 \\ \nu_4 \end{array} \right].
\end{equation}
In Eqs.~(6) to (10) we have consistently retained only the leading terms.

We now have our desired pattern of neutrino masses and mixing.  We see that
$\nu_\mu$ and $\nu_\tau$ are pseudo-Dirac partners with mass difference
squared given by
\begin{equation}
\delta m^2_{34} = {{2 m_7^2 m_2 m_3} \over {M_1 M_2}}.
\end{equation}
The singlet neutrino $\nu_S$ is mostly $\nu_2$ which has a small mass,
whereas $\nu_e$ is mostly $\nu_1$ which is massless.  Their mixing is
given by $m_1/m_4$.  Furthermore, $\nu_\mu$ oscillates to $\nu_e$ with
mixing given by $m_6/m_2$ as shown by Eq.~(10).  For illustration, let
$M_1 = M_2 = 100$ TeV, $m_2 = 10$ MeV, $m_3 = 25$ MeV, $m_4 = 0.5$ MeV,
$m_6 = 0.4$ MeV,
$m_7 = 0.6$ MeV, and $m_1 = 20$ keV; then the common mass of $\nu_\mu$
and $\nu_\tau$ is $m_2 m_3/M_2 = 2.5$ eV, the mass of $\nu_S$
is $m_4^2/M_1 = 2.5 \times 10^{-3}$ eV, $\delta m^2_{34} = 1.8 \times 10^{-2}$
eV$^2$, $\nu_e - \nu_S$ mixing is $m_1/m_4 = 0.04$, $\nu_e - \nu_\mu$
mixing is $m_6/m_2 = 0.04$, yielding an effective $\sin^2 2 \theta =
6.4 \times 10^{-3}$ in
either case as indicated by the solar-neutrino and LSND data.

With four light neutrinos, the nucleosynthesis bound\cite{17} of $N_\nu < 3.3$
is an important constraint.  Although $\nu_S$ is a singlet neutrino, it mixes
with $\nu_e$ and may contribute significantly to $N_\nu$ through
oscillations.  However, for the matter-enhanced nonadiabatic
$\nu_e - \nu_S$ oscillations which
explain the solar data, this is not a problem\cite{18}.  Note that if we had
made $m_1 >> m_4$ instead, then $\nu_S$ would be mostly massless and the
resulting scenario would be excluded by nucleosynthesis.  By the same token,
if we had tried to let $\nu_\mu$ oscillate into $\nu_S$ to explain the
atmospheric neutrino data, it would also be in conflict with nucleosynthesis.
These requirements are sufficient to pin down uniquely our model of four
light neutrinos provided we restrict ourselves to only two-oscillator
effects\cite{8} and the viewpoint that two almost degenerate neutrinos should
be pseudo-Dirac.  Of course, this forces us to have maximal mixing in the
$\nu_\mu - \nu_\tau$ sector and we must disregard part of the Frejus
data\cite{19}, but if a smaller effective $\sin^2 2 \theta$
(say 0.7) is desired, then the underlying symmetry as well as the mechanism
for generating ${\cal M}'_\nu$ become much more complicated\cite{7}.  As it is,
all we need is a discrete $Z_5$ symmetry and the seesaw mechanism.

Although $\nu_e - \nu_\tau$ mixing is very small as given by Eq.~(10), it
may instead come from the charged-lepton mass matrix of Eq.~(1).  Hence
our model can also accommodate such an effect.  Because two different
Higgs doublets contribute to the lepton mass matrices, flavor-changing
neutral-current processes do occur via the exchange of scalar bosons.
For example, $\mu \rightarrow e \gamma$ is possible, but because all the
Yukawa couplings are suppressed in this model, these are all negligible
as far as present experimental bounds are concerned.

Regarding the Higgs sector which consists of two doublets and a singlet,
it can be shown that the imposition of our discrete $Z_5$ symmetry
actually results in a continuous U(1) symmetry which is of course
broken in the Yukawa sector as $\chi^0$ couples both to $\nu_{eR} \nu_{SR}$
and $\nu_{\tau R} \nu_{\tau R}$.  This means that a pseudo-Goldstone boson
appears with a mass of order $m_5/4 \pi$ which is too small to be
compatible with present data.  To avoid this problem, a simple solution
is to enlarge the scalar sector with a real singlet $\eta_0$ and a neutral
complex singlet $\eta_2$ transforming under $Z_5$ as 1 and $\omega^2$
respectively.  In order that they do not couple directly to the leptons,
they are also assumed to be odd under an extra discrete $Z_2$ symmetry.
Because of the newly allowed terms $\eta_2 \eta_2 \chi$, $\eta_0 \eta_2
\bar \chi \bar \chi$, and $\eta_0 \eta_0$, there is no longer any unwanted U(1)
symmetry in this case and all the scalar bosons can be heavy.

The $4 \times 4$ neutrino mass matrix given by Eq.~(4) is similar but not
identical to those of Ref.~[7].  We assume that $\nu_\mu$ and $\nu_\tau$
to be pseudo-Dirac partners whereas Peltoniemi and Valle took $\nu_\mu$
and a linear combination of $\nu_\tau$ and $\nu_e$, while Caldwell and
Mohapatra did not consider them as pseudo-Dirac partners at all.  Our
model also differs in that we have a massless eigenvalue in the neutrino
mass matrix and they do not.  However, the most important difference
is in how we realize the desired form of the mass matrix.  We use a simple
discrete $Z_5$ symmetry and the seesaw mechanism whereas both papers of
Ref.~[7] require more complicated symmetries as well as radiative mechanisms
for mass generation.  Hence their scalar sectors are much more involved
and contain many particles of exotic hypercharge whereas we have only the
usual doublets
and neutral singlets.  Presumably, these are experimentally accessible at the
electroweak energy scale and would serve as discriminants of one model
against another.

In conclusion, we have shown that with all present desiderata, a simple
model of four light neutrinos can be constructed with pairwise oscillations
explaining the solar neutrino deficit ($\nu_e - \nu_S$), the atmospheric
neutrino anomaly ($\nu_\mu - \nu_\tau$), and the LSND observation ($\nu_\mu
- \nu_e$).  If the $\delta m^2$ of the last experiment is indeed 6 eV$^2$,
then there is also the cosmological connection of neutrinos as candidates
for hot dark matter, though that would be in marginal conflict
with another neutrino experiment\cite{20}.  If the experimental
inputs chosen by us do indeed stand the test of time, the matrix of Eq.~(2)
with the hierarchy of Eq.~(3) could well be the harbinger of a complete
theory of neutrino masses and mixing.
\vspace{0.5in}

\begin{center} {ACKNOWLEDGEMENT}
\end{center}

We thank Marc Sher for his helpful comments.  One of us (P.R.) acknowledges
the hospitality of the Physics Department of the University of California,
Riverside for an extended visit.
This work was supported in part by the U.S. Department of Energy under
Grant No. DE-FG03-94ER40837.

\newpage
\bibliographystyle{unsrt}

\end{document}